\documentclass[aps,prd,nofootinbib,showkeys,preprint,floatfix]{revtex4}

\usepackage{amssymb}
\usepackage{amsmath}
\usepackage{amsfonts}
\usepackage{graphicx}
\usepackage{color}
\usepackage{xspace}
\usepackage{ulem}
\usepackage{lscape}
\usepackage{ wasysym }

\usepackage{multirow,rotating}

\def\gsim{\raise0.3ex\hbox{$\;>$\kern-0.75em\raise-1.1ex\hbox{$\sim\;$}}}
\def\lsim{\raise0.3ex\hbox{$\;<$\kern-0.75em\raise-1.1ex\hbox{$\sim\;$}}}

\def\znbb{0\nu\beta\beta}

\newcommand{\ba}[1]{\begin{eqnarray} \label{(#1)}}
\newcommand{\ea}{\end{eqnarray}}

\newcommand{\AddrAHEP}{
  {\it AHEP Group, Instituto de F\'{\i}sica Corpuscular --
    C.S.I.C./Universitat de Val{\`e}ncia \\
    Edificio Institutos de Investigacion, Parc Cientific de Paterna, 
  Apartado 22085,
  E--46071 Val{\`e}ncia, Spain}}

\newcommand{\AddrUFSM}{
Universidad T\'ecnica Federico Santa Mar\'\i a, \\ 
Centro-Cient\'\i fico-Tecnol\'{o}gico de Valpara\'\i so, \\ 
Casilla 110-V, Valpara\'\i so,  Chile}

\def\gsim{\raise0.3ex\hbox{$\;>$\kern-0.75em\raise-1.1ex\hbox{$\sim\;$}}}
\def\lsim{\raise0.3ex\hbox{$\;<$\kern-0.75em\raise-1.1ex\hbox{$\sim\;$}}}

\begin{document}

\preprint{IFIC/15-51}

\title{LHC dijet constraints on double beta decay}

\author{J.C. Helo} \email{juancarlos.helo@usm.cl}\affiliation{\AddrUFSM}
\author{M. Hirsch} \email{mahirsch@ific.uv.es}\affiliation{\AddrAHEP}



\begin{abstract}

We use LHC dijet data to derive constraints on neutrinoless double
beta decay. Upper limits on cross sections for the production of
``exotic'' resonances, such as a right-handed W boson or a diquark,
can be converted into lower limits on the double beta decay half-life
for fixed choices of other parameters. Constraints derived from run-I
data are already surprisingly strong and complementary to results from
searches using same-sign dileptons plus jets.  For the case of the
left-right symmetric model, in case no new resonance is found in
future runs of the LHC and assuming $g_L=g_R$, we estimate a lower
limit on the double beta decay half-live larger than $10^{27}$ ys can
be derived from future dijet data, except in the window of relatively
light right-handed neutrino masses in the range $0.5$ MeV to $50$
GeV. Part of this mass window will be tested in the upcoming SHiP
experiment. We also discuss current and future limits on possible
scalar diquark contributions to double beta decay that can be 
derived from dijet data.

\end{abstract}

\maketitle


\section{Introduction}

Current experimental data on neutrinoless double beta decay ($\znbb$)
give limits for $^{76}$Ge \cite{Agostini:2013mzu} and $^{136}$Xe
\cite{Albert:2014awa,Shimizu:2014xxx,Gando:2012zm} in the range of
$T_{1/2}^{\znbb} \gsim (1-2) \times 10^{25}$ ys.  Proposals for next
generation $\znbb$ experiments even claim $T^{0\nu\beta\beta}_{1/2}
\sim 10^{27}$ yr can be reached for $^{136}$Xe
\cite{KamLANDZen:2012aa,Auty:2013:zz} and $^{76}$Ge
\cite{Abt:2004yk,Guiseppe:2011me}. Usually, these limits are
interpreted in terms of upper limits on Majorana neutrino masses.
However, any lepton number violating extension of the standard model
will contribute to $\znbb$ decay at some level and exchange of some
TeV-scale exotic particles could give even the dominant contribution
to the total $\znbb$ decay rate, see for example the recent reviews
\cite{Deppisch:2012nb,Hirsch:2015cga}.

\begin{figure}
\centering
\includegraphics[scale=0.45]{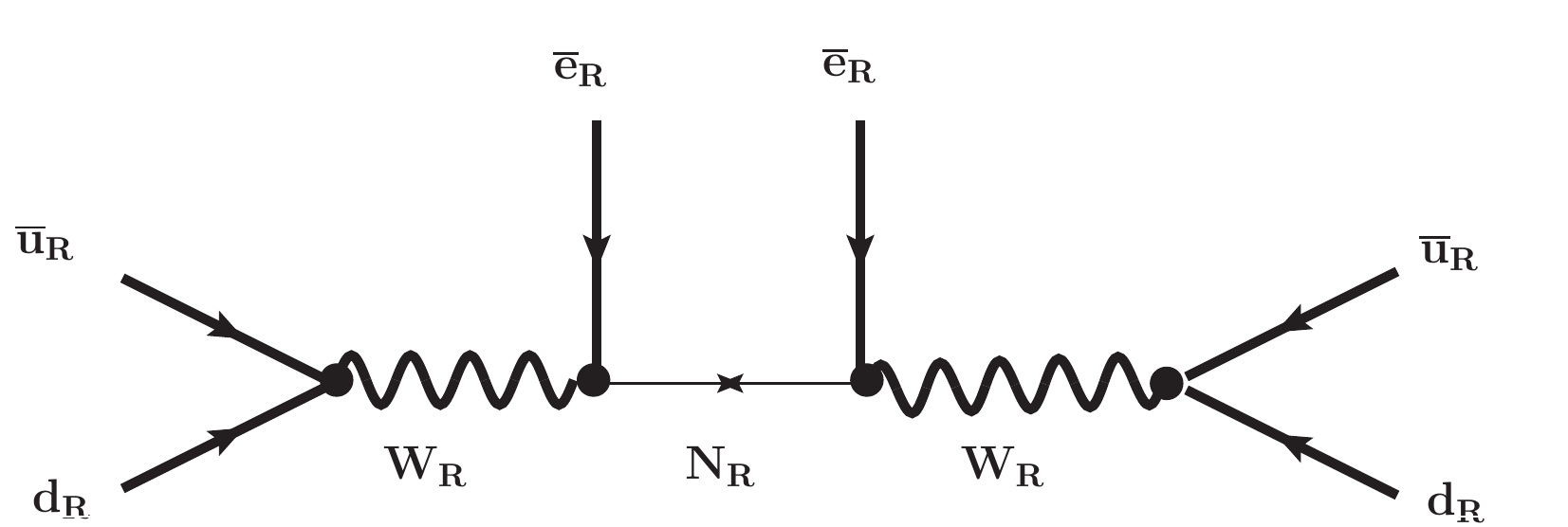}
\includegraphics[scale=0.55]{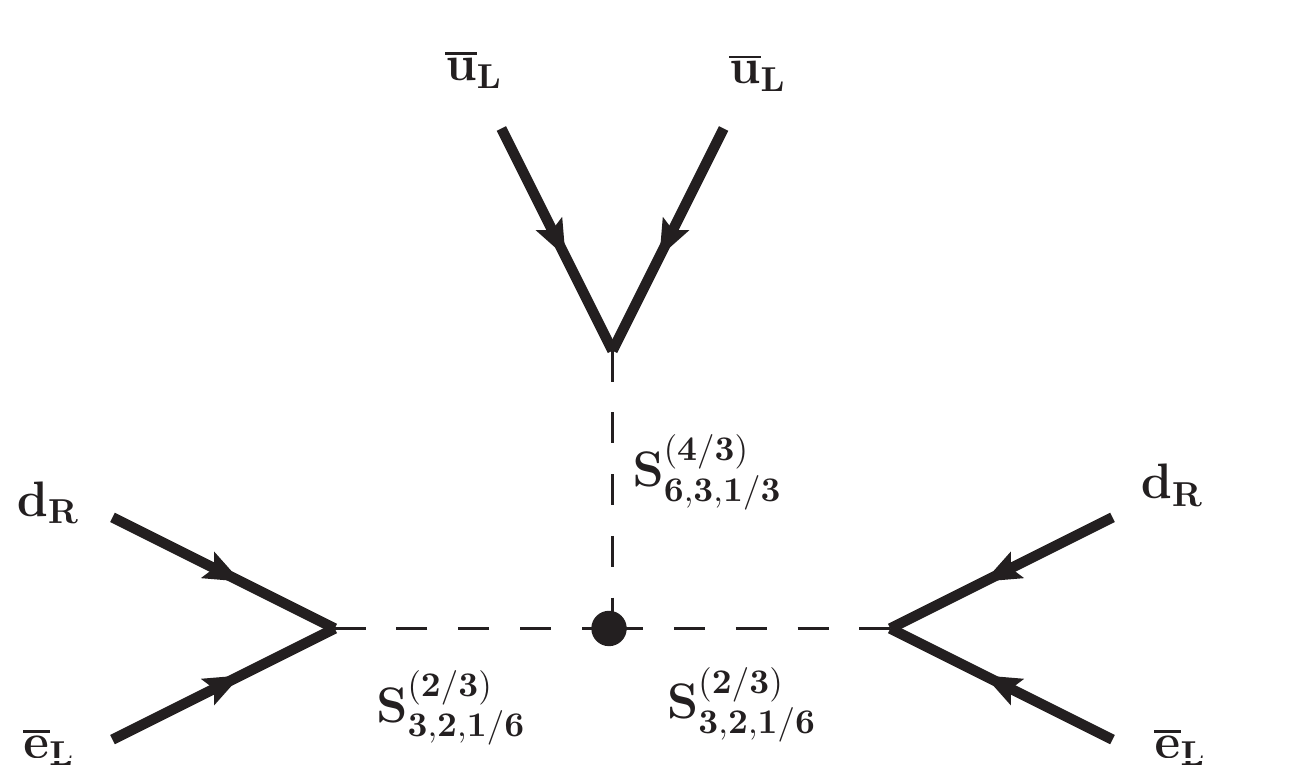}
\caption{\label{fig:diag}Two example diagrams of short-range
  contributions to double beta decay. To the left: (a) Left-right
  symmetric model, example of $W_R-N_R-W_R$ exchange (``topology-I''
  contribution); to the right: (b) a scalar diquark model classified
  as T-II-4 in \cite{Bonnet:2012kh} (``topology-II'' type
  contribution). For discussion see text.}
\end{figure}

The classical example of such a short-range contribution to $\znbb$
decay \cite{Pas:2000vn} is the right-handed W-boson exchange diagram
in left-right (LR) symmetric models \cite{Mohapatra:1980yp}, see
fig. (\ref{fig:diag}) to the left. Here, $N_{R_i}$ are the
right-handed partners of the ordinary neutrinos $\nu_{L_i}$. 
The general classification of all possible decompositions of the $d=9$
$\znbb$ decay operator can be found in \cite{Bonnet:2012kh}. In the
language of \cite{Bonnet:2012kh}, the diagram in fig. (\ref{fig:diag})
left is an example for a topology-I model. Topology-II contributions
to $\znbb$ decay, on the other hand, introduce no new fermions.  To
choose one particular example for T-II from the list of
\cite{Bonnet:2012kh} we take T-II-4, BL\#11.  Here, BL\# 11 refers to
operator ${\cal O}_{11}$ in the list of effective $\Delta L=2$
operators of \cite{Babu:2001ex}.  This model introduces a scalar
diquark, $S_{DQ}\equiv S_{6,3,1/3}$\footnote{Here and everywhere else
  in this paper subscripts denote the transformation properties/charge
  under the standard model gauge group, $SU(3)_c\times SU(2)_L\times
  U(1)_Y$.}, and a leptoquark, $S_{LQ}\equiv S_{3,2,1/6}$. The
short-range diagram contributing to $\znbb$ decay in this model is
shown in fig. (\ref{fig:diag}) on the right. We will come back to
discuss more details of diquarks in $\znbb$ decay in the next
section. Here, we only mention in passing that a possible $SU(5)$
embedding of this model has been recently discussed in
\cite{Fonseca:2015ena}.

At pp-colliders the classical signal of lepton number violation (LNV)
is the final state with two same-sign leptons plus two jets and no
missing energy ($lljj$). This signal was first discussed in the
context of left-right symmetric models in \cite{Keung:1983uu}, where
it can be simply understood as reading the diagram in
fig. (\ref{fig:diag}), from left to right.  Both, ATLAS
\cite{Aad:2015xaa} and CMS \cite{Khachatryan:2014dka} have searched
for this signal and give upper limits on $\sigma\times Br$ as a
function of the resonance mass. \footnote{In the CMS data there is an
  excess around $(2-2.2)$ TeV with a 2.8 $\sigma$ c.l. local
  significance. The ATLAS data, however, does not confirm this excess.
  We thus consider it a statistical fluctuation.}  These limits can
then be converted into excluded regions in parameter space for
different models.  For the example of the left-right symmetric model,
for right-handed neutrino masses, $m_{N_R}$, of the order of $m_{N_R}
\simeq \frac{1}{2} m_{W_R}$, this leads to very strong lower limits on
$ m_{W_R}$ of the order of $m_{W_R} \gsim (2.7-3)$ TeV
\cite{Aad:2015xaa,Khachatryan:2014dka}, assuming $g_R=g_L$. However,
these limits deteriorate rapidly if right handed neutrinos are
relatively light ($m_{N_R} \lsim 100$ GeV) or heavy ($m_{N_R} \gsim
m_{W_R} - 100$ GeV). In the former case, the lepton and the jets
emitted in the decay of $N_R$ are highly boosted and thus the lepton
is no longer isolated, failing one of the basic selection criteria
used by both LHC collaborations. If, on the other hand, $m_{N_R}$
approaches $m_{W_R}$, the jets and the lepton from the $N_R$-decay
become to soft to pass elementary $p_T$-cuts. Finally, for $m_{N_R}
\gsim m_{W_R}$, $N_R$ contributes only off-shell to the decay of $W_R$
and the branching ratio for the decay $W_R\to lljj$ drops to
unmeasurably small values.

One can use ATLAS \cite{Aad:2015xaa} and CMS
\cite{Khachatryan:2014dka} limits to constrain also all other models
with short-range contributions to the $\znbb$ decay rate. Diquarks
have particularly large cross sections at the LHC \cite{Han:2010rf},
so in the kinematic region where $m_{S_{LQ}} < m_{S_{DQ}}/2$
constraints from the $lljj$ search can be expected to be even more
severe than for LR models. Contributions to $\znbb$ decay from
leptoquark models, on the other hand, are less constrained from LHC
data. In \cite{Helo:2013dla,Helo:2013ika} current limits and expected
sensitivities based on the $lljj$ search for run-II have been
discussed for all T-I decompositions in the list of
\cite{Bonnet:2012kh}.

ATLAS \cite{Aad:2014aqa} and CMS \cite{Khachatryan:2015sja} have
searched for heavy, narrow resonances decaying to pairs of jets. No
clear signal for any new state has been found and both collaborations
provide upper limits on production cross sections times branching
ratio as function of the unknown resonance mass. These limits can be
converted into an upper limit on the unknown coupling of the resonance
to quarks (or, less interesting for us: gluons) as a function of the
resonance mass. In this paper, we discuss how these limits can be used
to constrain short-range contributions to $\znbb$ decay, despite the
fact that no LNV is searched for in the dijet data. As we discuss
below, the limits we derive are complementary to the limits derived
from the $lljj$ search and are surprisingly strong already with only
run-I data. We also estimate future LHC sensitivities and their
implications for $\znbb$ decay. In our numerical analysis, we
concentrate on the two example models, shown in fig. (\ref{fig:diag}), 
but also comment briefly on other possible contributions to 
$\znbb$ decay. 

The rest of this paper is organized as follows. In section \ref{sect:basics} 
we repeat very briefly the basics of the two models, which we use 
as examples.  Section \ref{sect:num} gives our numerical results. 
We then close in section \ref{sect:sum} with a short discussion.

\section{Model basics}
\label{sect:basics}

Our general arguments will apply to any ($\Delta L=2$) model
containing an exotic scalar or vector, which couples to a pair of
quarks.  For definiteness, we use the following two examples: (1) The
minimal left-right symmetric model and (2) a scalar diquark model. In
this section we briefly recall the basics of these two setups.

\subsection{Left-right symmetry}

The minimal left-right symmetric model extends the standard model
gauge group to $SU(3)_C\times SU(2)_L \times SU(2)_R \times
U(1)_{B-L}$ \cite{Pati:1974yy,Mohapatra:1974gc,Mohapatra:1980yp} and
assigns both left- and right-handed fermion fields as doublets (under
left and right groups, respectively). Thus the model contains
necessarily three generations of right-handed neutrinos. Charged and
neutral current interactions of the new gauge bosons are given by
\begin{eqnarray}\label{CC-NC-LR}
{\cal L} &=& \frac{g_{R}}{\sqrt{2}}  
 \left( V^{R}_{ud}\cdot\bar{d}  \gamma^{\mu} P_R u 
+  V^{R}_{l N}\cdot \bar l \gamma^{\mu} P_R N\  \right) W^-_{R \mu} 
 \\
\nonumber
&+&\frac{g_{R}}{\sqrt{1-\tan^{2}\theta_{W} (g_{L}/g_{R})^{2}}}
Z_{LR}^{\mu}\bar{f} \gamma_{\mu}
\left[ T_{3R} + \tan^{2}\theta_{W} (g_{L}/g_{R})^{2}\left(T_{3L}-Q\right)\right]f,  
\end{eqnarray}
Here, $V^{R}_{l N}$ ($ V^{R}_{ud}$) is the right-handed sector lepton
(quark) mixing, $g_L$, $g_R$ are the gauge couplings and $\theta_W$ is
the Weinberg angle. Eq.(\ref{CC-NC-LR}) shows that the couplings of
the $Z'$ boson to fermions becomes non-perturbative, if $g_R \lsim g_L
\tan\theta_W \simeq 0.35$. Very often in the literature it is assumed
that $g_R = g_L$, a special case which we will call manifest
left-right symmetry. However, in the numerical section we will 
allow $g_R$ also to vary.

In the minimal LR model, Majorana masses for the right-handed
neutrinos are generated by the vacuum expectation value breaking the
$SU(2)_R \times U(1)_{B-L}$ symmetry. One thus expects naively that
the $m_{N_{R_i}}$ are of the same order as the right-handed W-boson
mass, albeit times an unknown Yukawa coupling. To be as general as
possible, however, we will let these masses float freely. The
half-life $T_{1/2}$ for $\znbb$ decay via heavy $W_R$ and heavy $N_i$
exchange can then be written as:
\begin{eqnarray}\label{NLDBD-LR}
T^{-1}_{1/2} =  G_{01} 
\left|\sum_i
 \left(V^{R}_{eN_i}\right)^{2} m_{N_{R_i}}  {\cal{M}}(m_{N_{R_i}})   
\times \frac{m^{4}_{W_{L}}}{m^{4}_{W_{R}}} \frac{g^{4}_{R}}{g^{4}_{L}}
\right|^{2}
\end{eqnarray} 
Here, ${\cal{M}}(m_{N_{R_i}})$ is a nuclear matrix element, which depends
on $m_{N_{R_i}}$, and $G_{01}$ is the leptonic phase space integral. We
will use the numerical values of \cite{Hirsch:1996qw} for
${\cal{M}}(m_{N_{R_i}})$ in our analysis. For $m_{N_{R_i}}$ larger than
approximately $p_F \simeq {\cal O}(0.1-0.2)$ GeV, ${\cal{M}}(m_{N_{R_i}})
\propto \frac{1}{m_{N_{R_i}}^2}$ and we define the ``effective
right-handed neutrino mass'' as
\begin{eqnarray}\label{eq:meffR}
\frac{1}{\langle m_{N} \rangle}  = \sum_i
 \left(V^{R}_{eN_i}\right)^{2}  \frac{1}{m_{N_{R_i}}}.  
\end{eqnarray}
Note that due to the presence of Majorana phases 
there can be cancellations among terms in $\langle m_{N} \rangle$, 
which could lead to vastly larger values of the half-live but 
never to a shorter one compared to the case without Majorana phases. 
The latter is important, when deriving lower limits on $T_{1/2}$ from 
LHC data. 

For our analysis the exact fit to neutrino oscillation data is
unimportant. However, for completeness we mention that the minimal LR
model can explain this data at tree-level via the seesaw mechanism
\cite{Minkowski:1977sc,Yanagida:1979as,GellMann:1980vs,Mohapatra:1979ia}.
Naive expectation gives heavy-light neutrino mixing in the seesaw as
$V \propto \sqrt{m_{\nu}/M_N}$, i.e. $|V_{l 4}|^2 \simeq 5 \times
10^{-14} (\frac{m_{\nu}}{\rm 0.05 eV})(\frac{\rm 1 TeV}{M_{N_R}})$. Thus,
barring immensely huge cancellations among different contributions to
$m_{\nu}$, for an ordinary seesaw in LR models one expects that $N$
decays through a $W_R$ to $l^{\pm}jj$, with nearly equal rates in
$l^+$ and $l^-$, with a branching ratio close to 100 \%.

\subsection{Scalar diquarks}

As the second example model we discuss scalar diquarks. We define
scalar diquarks as particles coupling to a pair of same-type
quarks. They can be either colour triplets or sextets.  In the
context of $\znbb$ decay, diquark contributions were first discussed
in \cite{Gu:2011ak,Kohda:2012sr}. A systematic list of all (scalar)
diquark contributions to $\znbb$ decay was given in
\cite{Bonnet:2012kh}. We will concentrate on one particular diquark
model for definiteness. Constraints on other models will be very
similar; we will comment briefly in the numerical section.

From the list of possible diquark decompositions \cite{Bonnet:2012kh},
we choose the example T-II-4, BL\# 11. This particular case introduces
a diquark $S_{DQ}\equiv S_{6,3,1/3}$ plus a leptoquark $S_{LQ}\equiv
S_{3,2,1/6}$, see fig. (\ref{fig:diag}). The Lagrangian of the model
can be written as
\begin{eqnarray}\label{Lag-S6}
{\cal L}_{DQ LQ }&=& {\cal L}_{SM}  + 
  g_1 \bar{Q} \cdot \hat{S}_{DQ} \cdot Q^{C}   
 + g_2 \bar{L} \cdot S_{LQ}^{\dagger} d_R 
+  \mu S_{LQ}S_{LQ}S_{DQ}^{\dagger} + {\rm h.c.}
 \end{eqnarray}
For convenience we introduced the notation $\hat{S}_{DQ}= S_{DQ,
  a}^{(6)} (T_{\bar{\bf 6}})^{a}_{IJ}$, with $I,J=1-3$ and $a=1-6$ the
color triplet and sextet indexes, respectively.  The symmetric
$3\times 3$ matrices $T_{\bf 6}$ and $T_{\bar{\bf 6}}$ can be found in
ref.~\cite{Bonnet:2012kh}. $g_1$ and $g_2$ are dimensionless Yukawas,
we suppress generation indices for brevity. $\mu$ has dimension of
mass. Note that the Lagrangian in eq. (\ref{Lag-S6}) necessarily
violates lepton number by two units. 

$\znbb$ decay is generated via the diagram in fig. (\ref{fig:diag}),
to the right. Since neither diquarks nor leptoquarks can have 
masses light compared to the nuclear Fermi scale, this diagram 
is always of the short-range type. The inverse half-life is then 
\begin{eqnarray}\label{NLDBD-DQ}
T^{-1}_{1/2} =  G_{01} \left|\epsilon_{DQ} {\cal{M}}_{DQ} \right|^2,
\end{eqnarray} 
where \cite{Bonnet:2012kh}
\begin{equation}\label{eq:MDQ}
{\cal{M}}_{DQ} = \frac{1}{48} {\cal{M}}_{1} -  \frac{1}{192}  {\cal{M}}_{2}
\end{equation}
with ${\cal{M}}_{1,2}$ as defined in  \cite{Pas:2000vn}, where 
numerical values for $^{76}$Ge can be found, for other isotopes 
see \cite{Deppisch:2012nb}. $\epsilon_{DQ}$ is given by
\begin{equation}\label{eq:eps}
\epsilon_{DQ} = \frac{2 m_p}{G_F^2}\frac{g_1g_2^2\mu}{m_{DQ}^2m_{LQ}^4}.
\end{equation}
%

The model under consideration does not contain any right-handed
neutrino, instead it generates neutrino masses at 2-loop order
\cite{Helo:2015fba}.  Since for our purposes the exact numerical fit
to neutrino data is not important, we will not discuss the details
here. See either \cite{Sierra:2014rxa} for a general discussion of
2-loop neutrino mass models and/or \cite{Kohda:2012sr}, where a very
similar diquark model (based on a down-type diquark) has been
discussed in more details.

\section{Numerical results}
\label{sect:num}

We use CalcHEP \cite{Pukhov:2004ca} to calculate the cross section for
$W_R$ production and MadGraph5 \cite{Alwall:2014hca} for the
calculation of the x-section of the diquark. We have checked against
existing results in the literature \cite{Han:2010rf} and found good
agreement. We will first discuss our results for the left-right
symmetric model.

For deriving the constraints we use the CMS \cite{Khachatryan:2015sja}
data. ATLAS \cite{Aad:2014aqa} data leads to very similar results.
Moreover, for estimating the future sensitivities, we make use of the
fit of the SM dijet distribution fitted to a Monte Carlo simulation as
given in ref. \cite{Richardson:2011df}.  We then have estimated future
limits coming from dijet searches for an assumed luminosity of 
${\cal L} = 300 \ \text{fb}^{-1} $.

\subsection{Left-right symmetric model}

The branching ratio of the decay of a $W_R$ boson into two jets can be
calculated as a function of its mass, once the masses of the
right-handed neutrinos are fixed. In our numerical calculation we take
into account decays of the $W_R$ to fermions. 
\footnote{Decays of the $W_R$ to SM bosons depend on the mixing angle
  between $W_R$ and SM $W$ boson \cite{Brehmer:2015cia}, which we
  assume is small for simplicity.}  For masses of $m_{N_{R_i}} >
m_{W_R}$ and $ m_t \ll m_{W_R}$, $Br(W_R \to j j)$ then reaches
approximately $Br(W_R \to j j) \simeq 2/3$.  For all $m_{N_{R_i}} \ll
m_{W_R}$, $Br(W_R \to j j)\simeq 1/2$ for $ m_t \ll m_{W_R}$.  Using
our calculated cross section $ \sigma(p p \to W_R) $, the $Br(W_R \to
j j) $ and the upper limits on production cross sections times
branching ratio from dijet searches at $\sqrt{s} = 8 \ \text{TeV}$ and
${\cal L} = 19.7 \ \text{fb}^{-1} $ given by
ref. \cite{Khachatryan:2015sja} we have then calculated limits for
$g_R$ as function of $m_{W_R}$ for the LR model. Upper limits ranging
from roughly $g_R \sim [0.25,\sqrt{4\pi}]$ result for $m_{W_R}$ in the
range $m_{W_R} \simeq [1.2,4.4]$ TeV.

\begin{figure}
\centering
\includegraphics[scale=0.61]{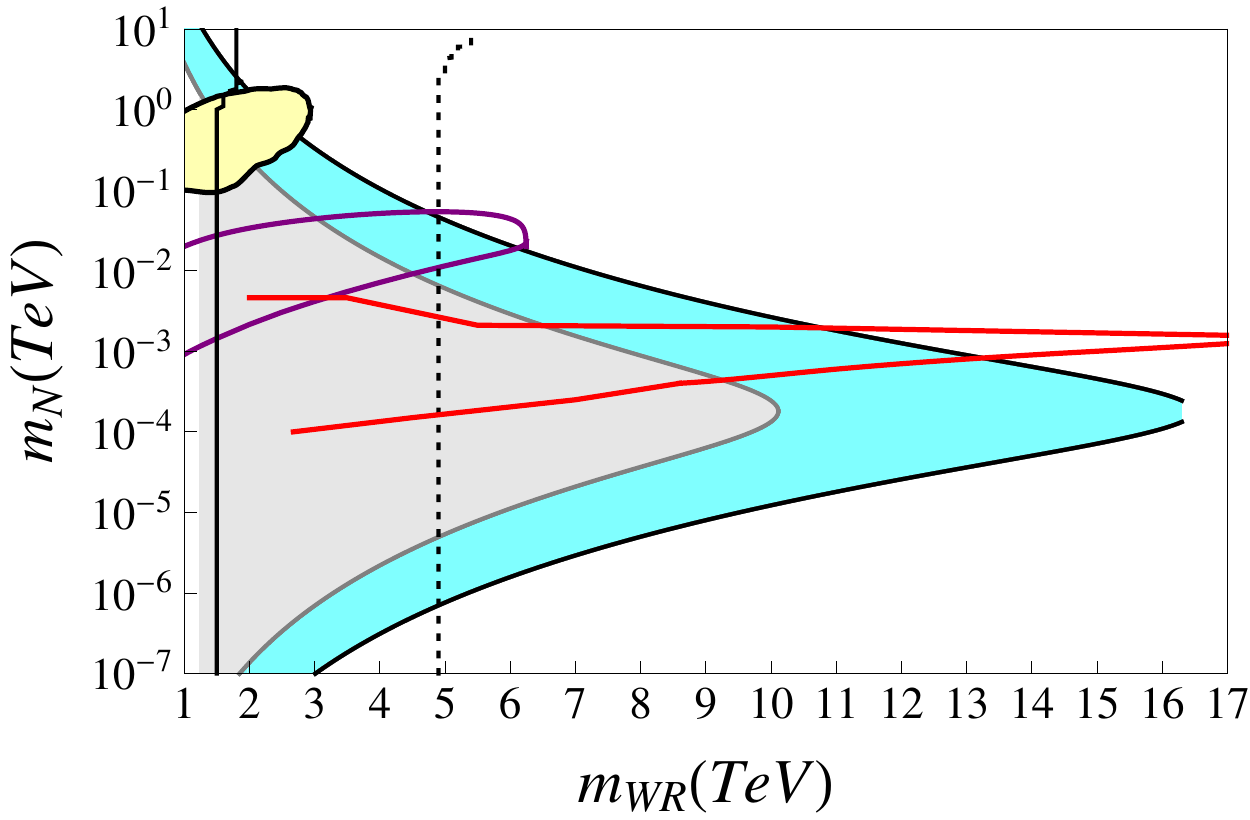}
\includegraphics[scale=0.61]{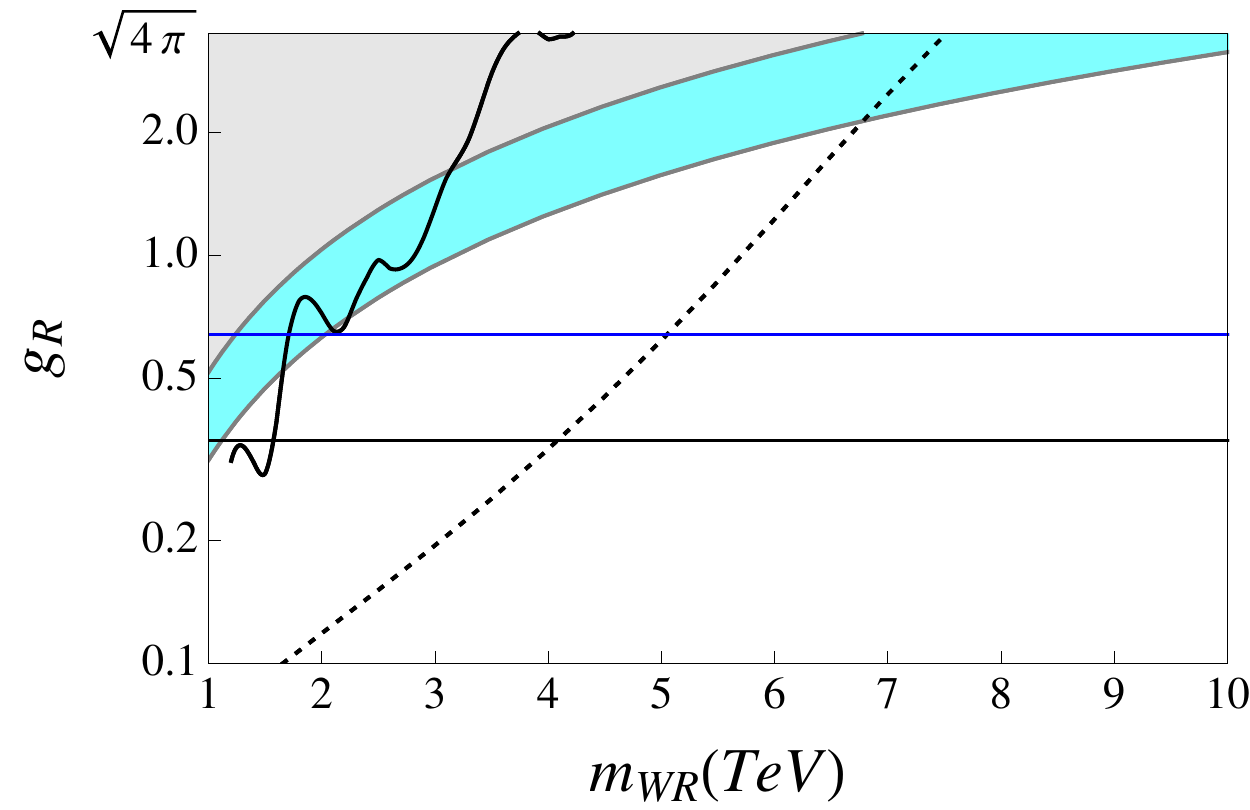}
\caption{\label{fig:gR}Regions in parameter space, which can be probed
  by dijet (black, full/dashed lines) and like sign leptons plus two
  jets (yellow region) searches at LHC, displaced vertex search at LHC
  (inside the purple lines), SHiP (red lines) and $\znbb$ decay. (a)
  Left: $ m_{W_R}$ vs $m_{N_{R}} $ for fixed $g_R = g_L$ (b) Right for
  fixed $m_N = 1\ \text{TeV}$ as a function of $g_R$. The black full
  (dashed) line are current (estimated future) LHC limits. The gray
  region is the current lower limit in $\znbb$ decay half-life of
  $^{76}$Ge, the blue one the estimated future sensitivity of $T_{1/2}
  = 10^{27}$ ys. For more details see text.}
\end{figure}

For the sake of simplicity consider first the case of manifest LR
symmetry, i.e., $g_R = g_L$, first.  In Fig. \ref{fig:gR}(a) we show
two limits from the non-observation of $\znbb$.  The gray region on
the left is ruled out by $\znbb$, corresponding to a half life
$T_{1/2} = 1.9 \times 10^{25} $ yr \cite{Albert:2014awa,
  Agostini:2013mzu}, while the stronger limit (blue region)
corresponds to an expected future sensitivity of $T_{1/2} = 10^{27}$
yr. Note that plots for $^{136}$Xe sensitivities are very similar.
The yellow region in the top corner shows CMS current limits from
searches of like-sign leptons plus two jets at $\sqrt{s} = 8
\ \text{TeV}$ and ${\cal L} = 19.7 \ \text{fb}^{-1} $
\cite{Khachatryan:2014dka}. Due to the choice of a logarithmic axes
for $m_N$, this region seems to represent only a tiny part of the
parameter space. However, we remind the reader that the naive
expectation for $m_N$ is typically $m_N \sim {\cal O}(m_{W_R})$.  The
solid red line in the middle of the plot shows the region in the
parameter space, which can be probed by heavy neutrino searches at the
upcoming SHiP experiment \cite{Alekhin:2015byh, Anelli:2015pba}. The
solid purple line shows the region in parameter space where a
displaced vertex search at the LHC could yield at least 5 events at
$\sqrt{s} = 13 \ \text{TeV}$ and ${\cal L} = 300 \ \text{fb}^{-1} $
\cite{Helo:2013esa, Castillo-Felisola:2015bha}.  The nearly vertical
solid (dotted) lines correspond to current (future) LHC limits from
dijet searched at $\sqrt{s} = 8 \ \text{TeV}$ ($13 \ \text{TeV}$) and
${\cal L} = 19.7 \ \text{fb}^{-1} $ ($300 \ \text{fb}^{-1} $). The
lines for the dijet limits assume three degenerate right-handed
neutrinos of mass $m_N$.  If only one right-handed has a mass below
$m_{W_R}$, the branching ratio Br($W_R\to jj$) increases, leading to
stronger limits from the dijet search.

Current searches of like-sign leptons plus two jets have imposed a
lower limit at $m_{W_R} \gsim (2.7 - 3.0)$ TeV in the neutrino mass
range $ 0.1 \ \text{TeV} \lesssim m_{N_{R}} \lesssim 2.0 \ \text{TeV}$
\footnote{These limits are expected to be extended up to $m_{W_R} = 5
  \ \text{TeV}$ in future searches \cite{Ferrari:2000sp}}. For this
part of the parameter region, the LHC limits rule out already a
dominant LR short-range contribution to $\znbb$. The current dijet
searches impose a lower limit at $m_{W_R} \simeq 1.5$ TeV ($2.0$ TeV)
for $\forall m_{N_{R_i}} < m_{W_R}$ ($\forall m_{N_{R_i}} > m_{W_R}$).
As can be seen from Fig. \ref{fig:gR}(a), dijet limits are
complementaries to those coming from like-sign leptons plus two jets,
extending the range also to the case $m_{N_R} > m_{W_R}$ and to 
$m_{N_R} \lsim 100$ GeV, although for such ``light'' right-handed 
neutrinos dijet searches are not yet competitive with $\znbb$ 
decay limits.  

Future dijets searches will impose strong limits at $m_{W_R} \lsim $ 5
TeV in case no new resonance is found at $13 \ \text{TeV}$ and ${\cal
  L} = 300 \ \text{fb}^{-1} $. As can be seen from
Fig. \ref{fig:gR}(a) these limits will leave only a small window for
LR short-range contribution for $\znbb$ experiments with half-lives of
order $10^{27}$ ($10^{25}$) ys at right-handed neutrino masses around
$ 1/2 \ \ \text{MeV} \lesssim m_{N_{R_i}} \lesssim 50 \ \text{GeV}$ ($
5 \ \ \text{MeV} \lesssim m_{N_{R_i}} \lesssim 7 \ \text{GeV}$). Part
of this window will be covered by SHiP (in the region of heavy
neutrino masses $m_{N_{R_i}} \sim 1-2 \ \text{GeV}$) and a possible
displaced vertex search \cite{Helo:2013esa,Castillo-Felisola:2015bha,
  Alekhin:2015byh}.

In Fig.~\ref{fig:gR}(b) we drop the assumption of manifiest LR
symmetry. Here we show, just as in Fig.~\ref{fig:gR}(a), a comparison
between the $0 \nu \beta \beta$ and dijet searches at LHC, but for
fixed heavy neutrino mass $m_{N_{R}} = 1 \ \text{TeV}$ in the plane
$g_R - m_{W_R}$. The blue horizontal line corresponds to the choice
$g_R = g_L$. The black horizontal line corresponds to the limit $g_R
\lsim g_L \tan\theta_W \simeq 0.35$ where the $Z'$ coupling to
fermions becomes non perturbative, as is explained in section
\ref{sect:basics}. As shown, and in agreement with the previous
analysis, dijet searches are competitive to $0 \nu \beta \beta$ for
$m_{N_{R}} = 1 \ \text{TeV}$, especially for small values of $g_R$.
For this choice of $m_{N_{R}}$, future $\znbb$ decay data can compete
with future LHC dijet data only for values of $g_R$ close to the
non-perturbative limit.

\begin{figure}
\centering
\includegraphics[scale=0.75]{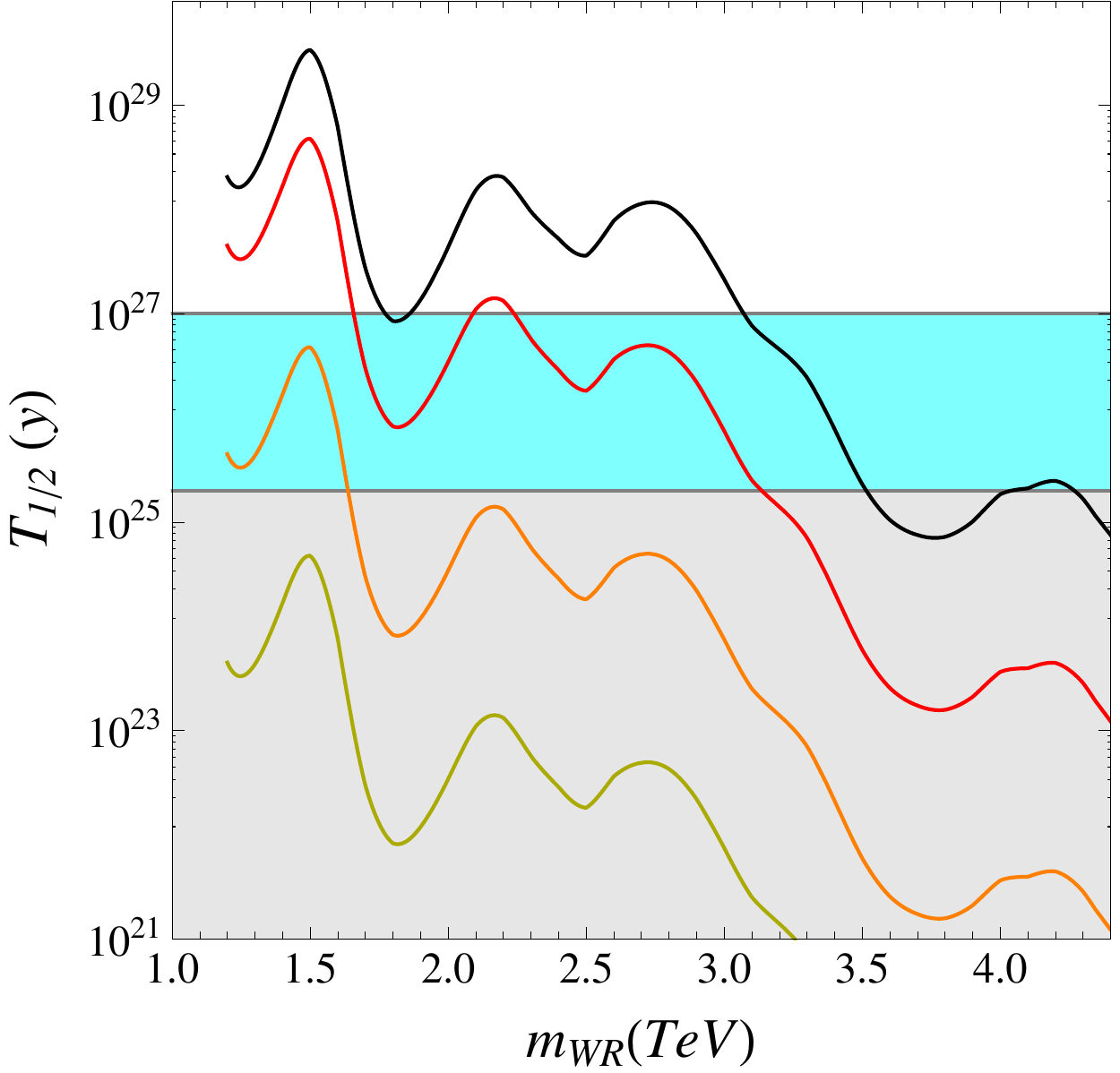}
\caption{\label{fig:limT12WR}Lower limit on the $\znbb$ decay
  half-life of $^{76}$Ge, derived from LHC dijet data for the
  left-right symmetric model for different values of the effective
  right-handed neutrino mass. The gray region is the current lower
  limit in $\znbb$ decay half-life of $^{76}$Ge, the blue one the
  estimated future sensitivity of $T_{1/2} = 10^{27}$ ys.  From top to
    bottom: $\langle m_N\rangle= m_{W_R}$, $\langle m_N\rangle=1$,
    $0.1$ and $0.01$ TeV. For $\langle m_N\rangle \ge m_{W_R}$
    half-lives below the experimental limit (straight line) are ruled
    out for values of $m_{W_R}$ up to $3.5$ TeV. For $m_{W_R} \simeq
    4.4$ TeV current LHC data do no longer give any constraint (the
    LHC limit on $g_R$ reaches $\sqrt{4 \pi}$).}
\end{figure}

Having fixed the limit on $g_R$ as function of $m_{W_R}$, we can then
calculate lower limits for half-lives for $\znbb$ decay, for different
assumed values of $\langle m_{N} \rangle$. Examples are shown for the
case of $^{76}$Ge in fig. (\ref{fig:limT12WR}) using current LHC dijet
limits. Note that the plots extend up to $m_{W_R} \simeq 4.4$ TeV; at
this point the limit on $g_R$ becomes worse than $g_R \simeq
\sqrt{4\pi}$, and the theory would be non-perturbative, i.e. the
limits no longer have any physical meaning. For $\langle m_{N} \rangle
= m_{W_R}$ the strongest lower limits result, for the whole region of
$m_{W_R}$ up to $m_{W_R} \simeq 3.5$ TeV half-life limits longer than
the current experimental limit can be derived. The constraints become
less stringent for smaller values of $\langle m_{N} \rangle$ and are
practically completely irrelevant for masses lower or equal than
$\langle m_N\rangle \simeq 10$ GeV using current LHC data. 

We close this discussion with a short comment on charged scalars.
Decompositions with singly charged scalars appear in the list of
short-range $\znbb$ decay contributions \cite{Bonnet:2012kh}.  The
cross section of $S_+$ at the LHC is typically around a factor of $2$
smaller than the cross section for a heavy $W'$, for the same value of
the coupling constants to quarks. Thus, similar albeit slightly weaker
limits, as discussed here for $W_R$, can be derived from dijet
searches also for charged scalar contributions to $\znbb$ decay.  One
slight complication arises for charged scalars, however, with respect
to the LR model discussed here: In a gauge model, like LR, the 
coupling of $W_R$ to quarks and leptons is universal, whereas for the 
charged scalar the couplings $g_{ud} ({\bar u}d)S_+$ and 
$g_{eN} ({\bar e}N)S_+^\dagger$ could in principle be different. 
If $g_{ud} \ne g_{eN} $ the discussion for charged scalars will 
resemble more the case of scalar diquarks, which we discuss next.

\subsection{Scalar diquark Model}

Now we turn to the results for the scalar diquark model.  As in the LR
symmetric case we have used the cross section $ \sigma(p p \to
S_{DQ})$ and the $Br(S_{DQ} \to j j)$ to calculate current and future
limits from dijet searches, using the upper limits on production cross
sections times branching ratio from dijet searches
\cite{Khachatryan:2015sja}. For estimating future sensitivities we use
the estimated QCD backgrounds from the Monte Carlo simulation
\cite{Richardson:2011df}.

For the sake of simplicity we will assume the Yukawa couplings $g_1$
and $g_2$ are different from zero for the first quark and lepton
generations only.  In this model the scalar diquark has two possible
decay modes: two jets $(jj)$ and two lepton plus two jets $(lljj)$. On
one hand, in the parameter region $m_{DQ} < 2 m_{LQ}$ the $Br(S_{DQ}
\rightarrow j j ) \simeq 1$ since $S_{LQ}$ contributes only off-shell
to the decay of $S_{DQ}\rightarrow S_{LQ}^* S_{LQ}^* \rightarrow lljj$
and the $Br(S_{DQ} \rightarrow l l j j)$ drops to unmeasurably small
values. On the other hand, in the region where $m_{LQ} << m_{DQ}$ the
$Br(S_{DQ} \rightarrow j j )$ becomes a function of also $m_{LQ}$ and
the (unknown) parameters $\mu$ and $g_2$, see eq.(\ref{Lag-S6}).

\begin{figure}
\centering
\includegraphics[scale=0.61]{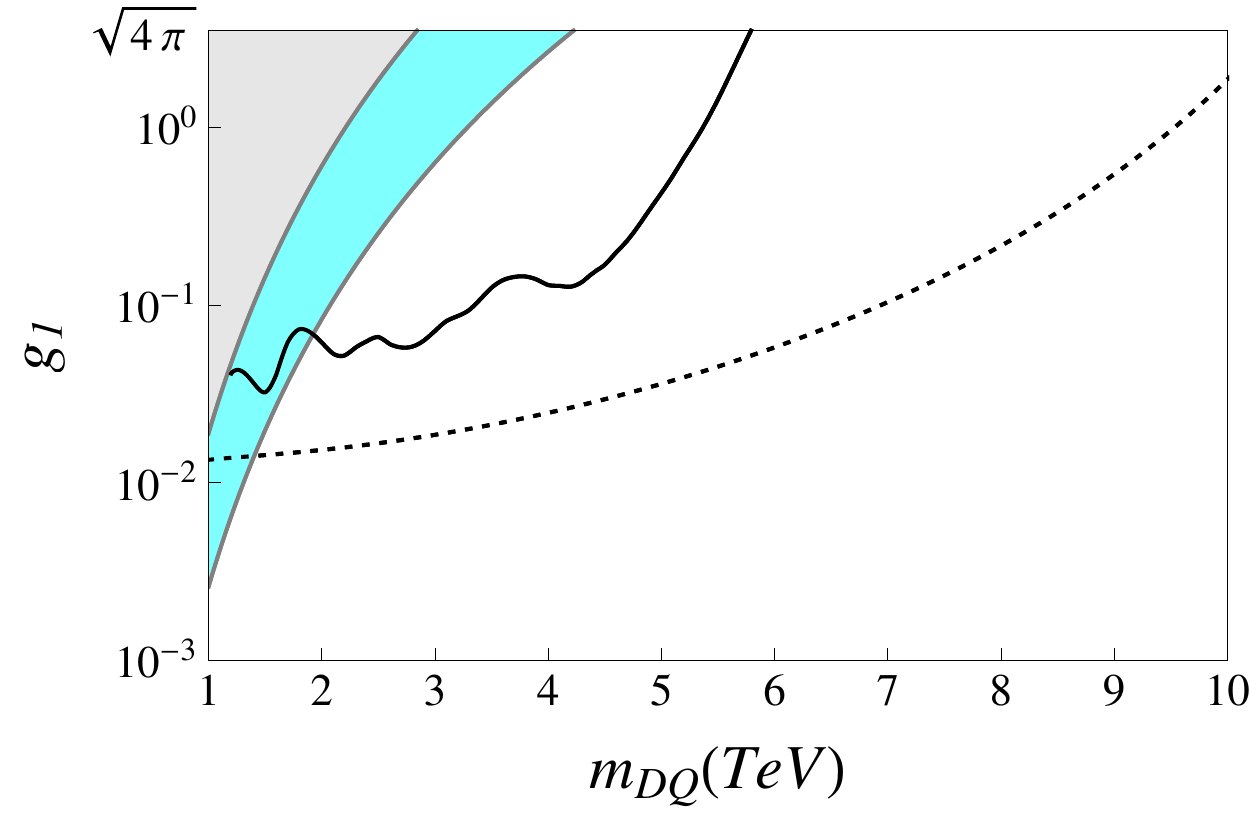}
\includegraphics[scale=0.61]{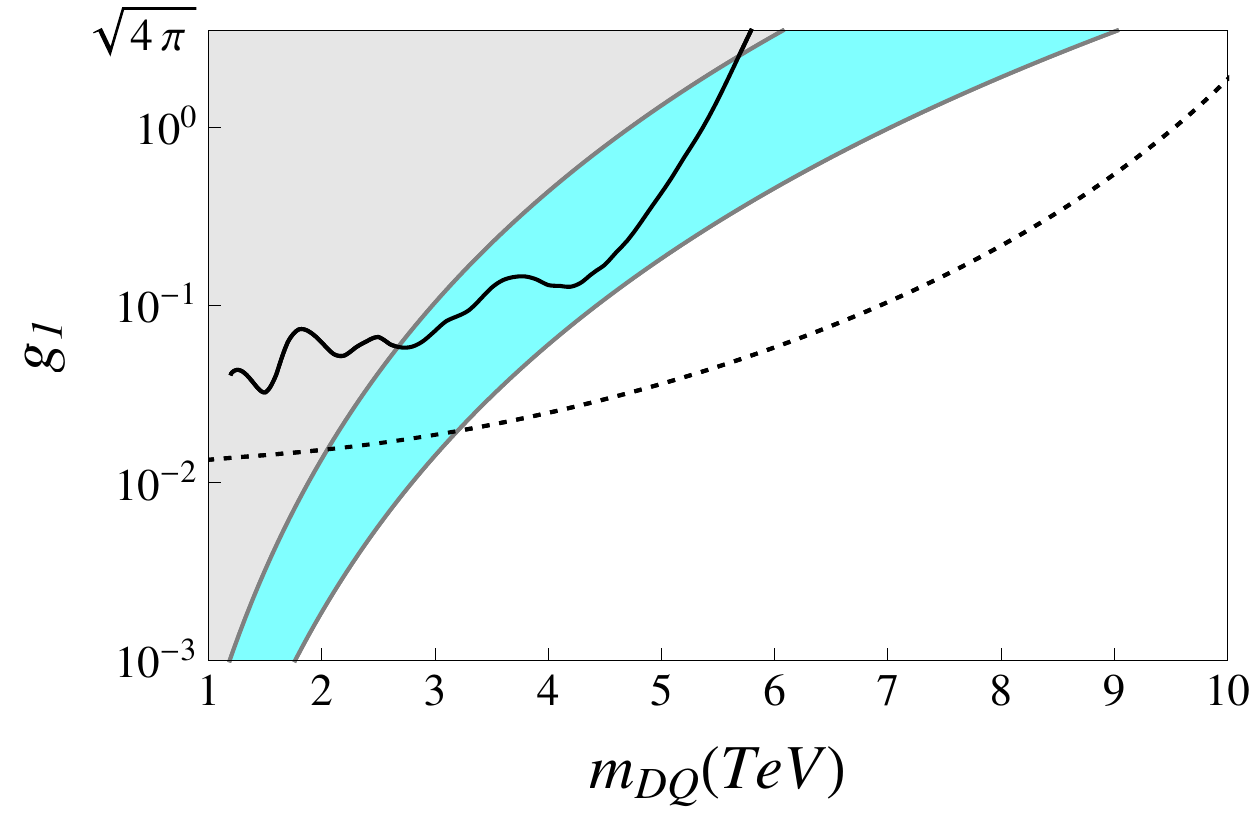}
\caption{\label{fig:g1} Future and current limits from dijet searches
  at LHC compared with double beta decay experiments. The gray region
  on the top left corner is ruled out by $\znbb$. The blue region 
  corresponds to an expected future sensitivity of $T_{1/2} = 10^{27}
  yr$. The solid (dotted) black line correspond to current (future)
  limits from dijet searches. The $\znbb$ limits were calculated using
  $m_{LQ} = m_{DQ}$, $g_{2}=1 \ \text{(left)}, \sqrt{4 \pi}
  \ \text{(right)}$ and $\mu = m_{DQ} \ \text{(left)}, \sqrt{4 \pi}
  \ m_{DQ} \ \text{(right)}$.}
\end{figure}

Consider first the simpler case $m_{LQ} \ge m_{DQ}/2$.  In
Fig. \ref{fig:g1} we show two limits from the non-observation of
$\znbb$.  The gray region on the left is ruled out by $\znbb$,
corresponding to a half life $T_{1/2} = 1.9 \times 10^{25} yr$
\cite{Albert:2014awa, Agostini:2013mzu}, while the stronger limit
(blue region) corresponds to an expected future sensitivity of
$T_{1/2} = 10^{27} yr$.  The solid (dotted) lines correspond to
current (future) LHC limits from dijet searched at $\sqrt{s} = 8
\ \text{TeV}$ ($13 \ \text{TeV}$) and ${\cal L} = 19.7
\ \text{fb}^{-1} $ ($300 \ \text{fb}^{-1} $).  Double beta decay
limits were calculated using, in Eq. (\ref{eq:eps}), $m_{LQ} = m_{DQ}
$, $\mu = m_{DQ}$, $g_2 =1 (\text{left})$ and $m_{LQ} = m_{DQ} $, $\mu
= \sqrt{4 \pi} \ m_{DQ}$, $g_2 =\sqrt{4 \pi} \ (\text{right})$. For
larger masses $m_{LQ} $ or smaller couplings $g_2$ and $\mu$ those
limits become weaker.  Note that the case $g_2 \equiv g_1$, which is
more similar to the case of the LR symmetric model, where uinversality
of couplings is enforced by the gauge symmetry, $\znbb$ sensitivities
would be much worse than the ones shown in this plot.  Already with
current LHC data, dijet limits are more stringent than current $\znbb$
decay limits in this part of parameter space, except for a window of
very small values of $g_1$ at small $m_{DQ}$. The large reach of the 
LHC simply reflects the large diquark production cross section.

\begin{figure}
\centering
\includegraphics[scale=0.61]{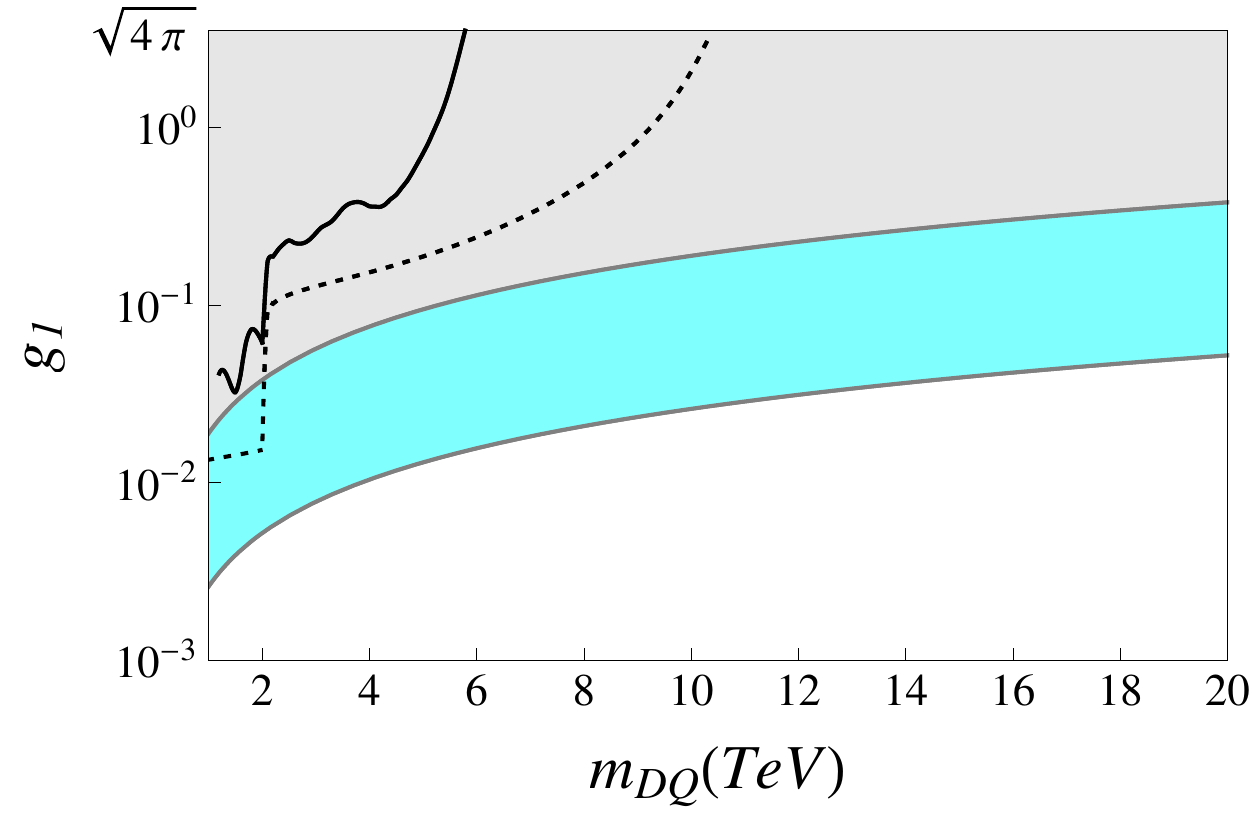}
\includegraphics[scale=0.61]{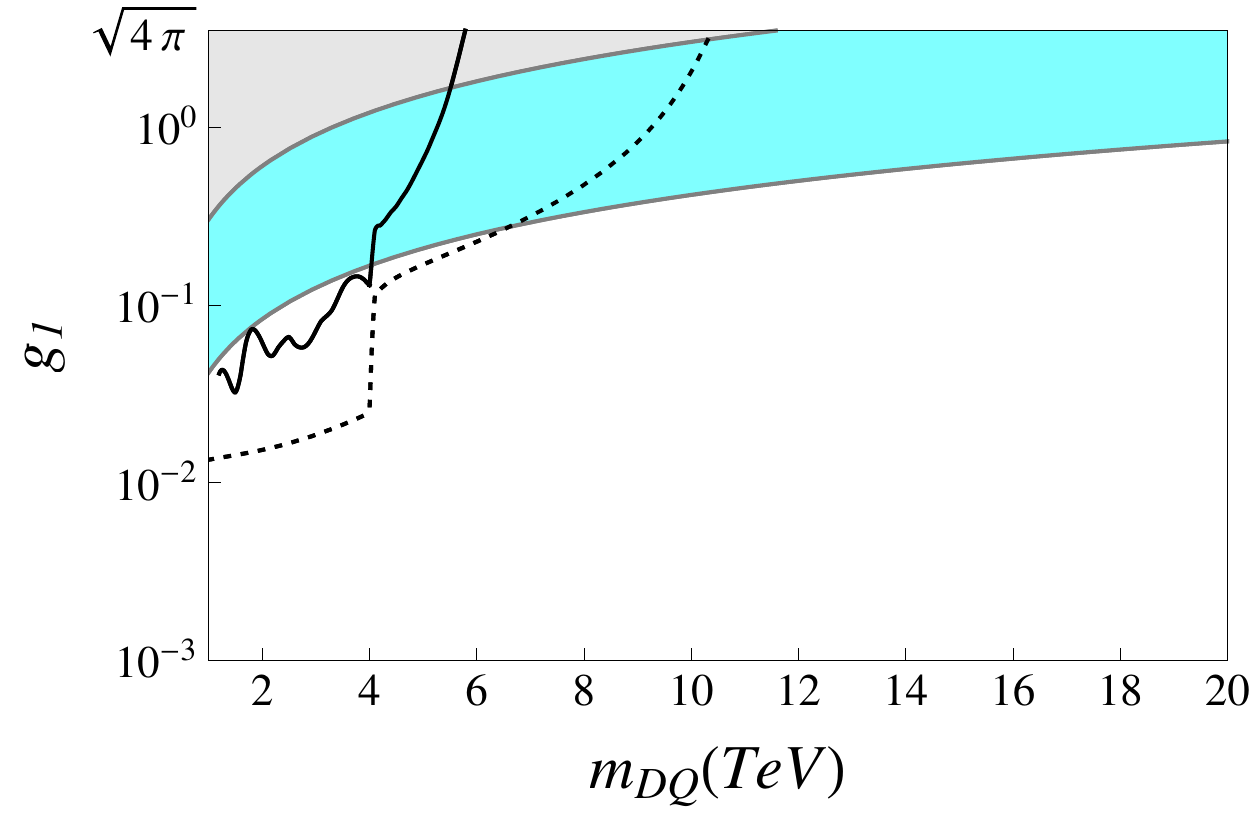}
\caption{\label{fig:g12} Future and current limits from dijet searches
  at LHC compared with double beta decay experiments. The gray region
  on the top left corner is ruled out by $\znbb$. The blue region 
  corresponds to an expected future sensitivity of $T_{1/2} = 10^{27}
  yr$. The solid (dotted) black line correspond to current (future)
  limits from dijet searches. The $\znbb$ and LHC limits were
  calculated using for $\mu = m_{DQ}$, $g_{2}=1$ and $m_{LQ} = 1 \ TeV
  \ \text{(left)}, 2 \ TeV \ \text{(right)}$.}
\end{figure}

In Fig.~\ref{fig:g12} we show, just as in Fig.~\ref{fig:g1}, a
comparison between the $0 \nu \beta \beta$ and dijet searches at LHC,
but for $\mu = m_{DQ}$, $g_2 =1$, $m_{LQ} = 1 \ \text{TeV (left), and}
\ m_{LQ}= 2 \ \text{TeV (right)}$. Smaller values of $m_{LQ}$ give
$\znbb$ decay a better sensitivity to $g_1$, while for these
relatively large values of $\mu$ the diquark has sizeable branching
ratio into $lljj$ final states, thus reducing the LHC sensitivity in
the dijet search. As fig.~\ref{fig:g12} shows, in this part of the
parameter space the dijet search can not fully compete with $\znbb$
decay. However, since this reduced sensitivity comes from the
competition between $lljj$ and $jj$ final states, one can expect that
this part of the parameter space can be covered with future lepton
number violating searches at the LHC. We plan to come back to study
this part of parameter space in more detail in a future publication on
topology-II $\znbb$ decay.

We close this section with a short comment on charged scalars and
other types of diquarks. Down-type diquarks have cross sections
roughly a factor $\sim (4-8)$ smaller than the up-type diquarks
discussed here, with charged scalars having smaller cross sections
still \cite{Helo:2013ika}. Thus, numerically weaker limits from dijet
searches are expected for these cases. However, the discussion for
these cases will be similar qualitatively. Therefore, we do not repeat
all details for charged scalars and down-type diquarks here.

\section{Summary}
\label{sect:sum}

We have discussed how upper limits on dijet cross sections, derived
from LHC data, can be used to constrain the short-range part of the
$\znbb$ decay amplitude. We have concentrated on two example models:
(a) minimal left-right symmetry and (b) a diquark model with LNV.  For
both setups, the LHC dijet data \cite{Aad:2014aqa,Khachatryan:2015sja}
provides constraints complementary to those derived from the search
for $lljj$ final state \cite{Aad:2015xaa,Khachatryan:2014dka}. We have
also estimated the impact of future LHC data. Current dijet limits
provide already interesting constraints on $\znbb$ decay, future
limits will rule out measurably ``small'' half-lives of double beta
decay ($T_{1/2} \lsim 10^{27}$ ys), except in some well-defined
regions of parameter space. The details for the different cases are
discussed in the main text. We note that, while we have concentrated
on two particular example models, similar constraints will apply to
any short-range contribution to $\znbb$ decay in which a state
coupling to a pair of quarks appars.

Finally, we note that dijet data can give interesting limits on
$\znbb$ decay, as long as no new physics is found in the search. If,
however, a new resonance were to appear in the data of run-II,
obviously dedicated $\Delta L=2$ searches will be needed to prove or
disprove any connection of such a hypothetical discovery to $\znbb$
decay. In this sense, dijet searches are complementary to the
``standard'' $lljj$ search at the LHC, but can not replace it 
as a discovery tool.

\bigskip
\centerline{\bf Acknowledgements}

\medskip

\medskip
M.H. is supported the Spanish grants FPA2014-58183-P, Multidark
CSD2009-00064 and SEV-2014-0398 (MINECO), and PROMETEOII/2014/084
(Generalitat Valenciana).
J.C.H. is supported by Fondecyt (Chile) under grant
11121557.



\begin{thebibliography}{10}

\bibitem{Agostini:2013mzu}
GERDA Collaboration, M.~Agostini {\em et~al.},
\newblock Phys.Rev.Lett. {\bf 111}, 122503 (2013), arXiv:1307.4720.

\bibitem{Albert:2014awa}
EXO-200 Collaboration, J.~Albert {\em et~al.},
\newblock Nature {\bf 510}, 229–234 (2014), arXiv:1402.6956.

\bibitem{Shimizu:2014xxx}
KamLAND-Zen Collaboration, I.~Shimizu,
\newblock Neutrino 2014, Boston  (2014).

\bibitem{Gando:2012zm}
KamLAND-Zen Collaboration, A.~Gando {\em et~al.},
\newblock Phys. Rev. Lett. {\bf 110}, 062502 (2013), arXiv:1211.3863.

\bibitem{KamLANDZen:2012aa}
KamLAND-Zen Collaboration, A.~Gando {\em et~al.},
\newblock Phys.Rev. {\bf C85}, 045504 (2012), arXiv:1201.4664.

\bibitem{Auty:2013:zz}
EXO-200 Collaboration, D.~Auty,
\newblock Recontres de Moriond, http://moriond.in2p3.fr/  (2013).

\bibitem{Abt:2004yk}
GERDA Collaboration, I.~Abt {\em et~al.},
\newblock (2004), arXiv:hep-ex/0404039.

\bibitem{Guiseppe:2011me}
Majorana Collaboration, C.~Aalseth {\em et~al.},
\newblock Nucl.Phys.Proc.Suppl. {\bf 217}, 44 (2011), arXiv:1101.0119.

\bibitem{Deppisch:2012nb}
F.~F. Deppisch, M.~Hirsch, and H.~P\"as,
\newblock J.Phys. {\bf G39}, 124007 (2012), arXiv:1208.0727.

\bibitem{Hirsch:2015cga}
M.~Hirsch,
\newblock AIP Conf. Proc. {\bf 1666}, 170007 (2015).

\bibitem{Pas:2000vn}
H.~P\"as, M.~Hirsch, H.~Klapdor-Kleingrothaus, and S.~Kovalenko,
\newblock Phys.Lett. {\bf B498}, 35 (2001), arXiv:hep-ph/0008182.

\bibitem{Mohapatra:1980yp}
R.~N. Mohapatra and G.~Senjanovic,
\newblock Phys. Rev. {\bf D23}, 165 (1981).

\bibitem{Bonnet:2012kh}
F.~Bonnet, M.~Hirsch, T.~Ota, and W.~Winter,
\newblock JHEP {\bf 1303}, 055 (2013), arXiv:1212.3045.

\bibitem{Babu:2001ex}
K.~Babu and C.~N. Leung,
\newblock Nucl.Phys. {\bf B619}, 667 (2001), arXiv:hep-ph/0106054.

\bibitem{Fonseca:2015ena}
R.~M. Fonseca and M.~Hirsch,
\newblock Phys. Rev. {\bf D92}, 015014 (2015), arXiv:1505.06121.

\bibitem{Keung:1983uu}
W.-Y. Keung and G.~Senjanovic,
\newblock Phys.Rev.Lett. {\bf 50}, 1427 (1983).

\bibitem{Aad:2015xaa}
ATLAS, G.~Aad {\em et~al.},
\newblock (2015), arXiv:1506.06020.

\bibitem{Khachatryan:2014dka}
CMS, V.~Khachatryan {\em et~al.},
\newblock Eur.Phys.J. {\bf C74}, 3149 (2014), arXiv:1407.3683.

\bibitem{Han:2010rf}
T.~Han, I.~Lewis, and Z.~Liu,
\newblock JHEP {\bf 1012}, 085 (2010), arXiv:1010.4309.

\bibitem{Helo:2013dla}
J.~Helo, M.~Hirsch, S.~Kovalenko, and H.~P\"as,
\newblock Phys.Rev. {\bf D88}, 011901 (2013), arXiv:1303.0899.

\bibitem{Helo:2013ika}
J.~Helo, M.~Hirsch, H.~P\"as, and S.~Kovalenko,
\newblock Phys.Rev. {\bf D88}, 073011 (2013), arXiv:1307.4849.

\bibitem{Aad:2014aqa}
ATLAS, G.~Aad {\em et~al.},
\newblock Phys.Rev. {\bf D91}, 052007 (2015), arXiv:1407.1376.

\bibitem{Khachatryan:2015sja}
CMS, V.~Khachatryan {\em et~al.},
\newblock Phys.Rev. {\bf D91}, 052009 (2015), arXiv:1501.04198.

\bibitem{Pati:1974yy}
J.~C. Pati and A.~Salam,
\newblock Phys.Rev. {\bf D10}, 275 (1974).

\bibitem{Mohapatra:1974gc}
R.~Mohapatra and J.~C. Pati,
\newblock Phys.Rev. {\bf D11}, 2558 (1975).

\bibitem{Hirsch:1996qw}
M.~Hirsch, H.~Klapdor-Kleingrothaus, and O.~Panella,
\newblock Phys.Lett. {\bf B374}, 7 (1996), arXiv:hep-ph/9602306.

\bibitem{Minkowski:1977sc}
P.~Minkowski,
\newblock Phys.Lett. {\bf B67}, 421 (1977).

\bibitem{Yanagida:1979as}
T.~Yanagida,
\newblock Conf.Proc. {\bf C7902131}, 95 (1979).

\bibitem{GellMann:1980vs}
M.~Gell-Mann, P.~Ramond, and R.~Slansky,
\newblock Conf.Proc. {\bf C790927}, 315 (1979),
\newblock Supergravity, P. van Nieuwenhuizen and D.Z. Freedman (eds.), North
  Holland Publ. Co., 1979.

\bibitem{Mohapatra:1979ia}
R.~N. Mohapatra and G.~Senjanovic,
\newblock Phys. Rev. Lett. {\bf 44}, 912 (1980).

\bibitem{Gu:2011ak}
P.-H. Gu,
\newblock Phys.Rev. {\bf D85}, 093016 (2012), arXiv:1101.5106.

\bibitem{Kohda:2012sr}
M.~Kohda, H.~Sugiyama, and K.~Tsumura,
\newblock Phys.Lett. {\bf B718}, 1436 (2013), arXiv:1210.5622.

\bibitem{Helo:2015fba}
J.~Helo, M.~Hirsch, T.~Ota, and F.~A.~P. Dos~Santos,
\newblock JHEP {\bf 1505}, 092 (2015), arXiv:1502.05188.

\bibitem{Sierra:2014rxa}
D.~Aristizabal~Sierra, A.~Degee, L.~Dorame, and M.~Hirsch,
\newblock JHEP {\bf 1503}, 040 (2015), arXiv:1411.7038.

\bibitem{Pukhov:2004ca}
A.~Pukhov,
\newblock (2004), arXiv:hep-ph/0412191.

\bibitem{Alwall:2014hca}
J.~Alwall {\em et~al.},
\newblock JHEP {\bf 1407}, 079 (2014), arXiv:1405.0301.

\bibitem{Richardson:2011df}
P.~Richardson and D.~Winn,
\newblock Eur. Phys. J. {\bf C72}, 1862 (2012), arXiv:1108.6154.

\bibitem{Brehmer:2015cia}
J.~Brehmer, J.~Hewett, J.~Kopp, T.~Rizzo, and J.~Tattersall,
\newblock (2015), arXiv:1507.00013.

\bibitem{Alekhin:2015byh}
S.~Alekhin {\em et~al.},
\newblock (2015), arXiv:1504.04855.

\bibitem{Anelli:2015pba}
SHiP, M.~Anelli {\em et~al.},
\newblock (2015), arXiv:1504.04956.

\bibitem{Helo:2013esa}
J.~C. Helo, M.~Hirsch, and S.~Kovalenko,
\newblock Phys. Rev. {\bf D89}, 073005 (2014), arXiv:1312.2900.

\bibitem{Castillo-Felisola:2015bha}
O.~Castillo-Felisola, C.~O. Dib, J.~C. Helo, S.~G. Kovalenko, and S.~E. Ortiz,
\newblock Phys. Rev. {\bf D92}, 013001 (2015), arXiv:1504.02489.

\bibitem{Ferrari:2000sp}
A.~Ferrari {\em et~al.},
\newblock Phys.Rev. {\bf D62}, 013001 (2000).

\end{thebibliography}
\end{document}